\documentstyle[twocolumn]{jpsj}

\title{Simulation of Spin Glass with a Variable-system-size Ensemble}

\author{ Yukito {\sc Iba}\footnote{e-mail: iba@ism.ac.jp}
\inst{The Institute of Statistical Mathematics, \\
4-6-7 Minami-Azabu Minato-ku , Tokyo, 106-8569, Japan } }

\recdate{\today}

\abst{ 
In this paper, we introduce a dynamical Monte Carlo
algorithm for spin models in which the number of the spins
fluctuates from zero to a given number 
by addition and deletion of spins with a probabilistic rule.
Such simulations are realized 
with a variable-system-size ensemble, a mixture of canonical ensembles
each of which corresponds to a system with different size. 
The weight of each component of the mixture
is controlled by a penalty term
and  systematically tuned  
in a preliminary run in a way similar to the multicanonical algorithm. 
In a measurement run, the system grows
and shrinks without violating the detailed balance condition and 
we can obtain the correct canonical averages
if physical quantities is measured only when 
its size is equal to the prescribed maximum size. 
The mixing of Markov chain 
is facilitated by the fast relaxation at small system sizes. 
The algorithm is implemented
for the SK model of spin glass and shows better performance
than that of a conventional heat bath algorithm.
}

\kword{Monte Carlo, extended ensemble, 
multicanonical, variable-system-size ensemble, relaxation, \mbox{spin glass} }

\begin{document}

\sloppy
\maketitle

In this paper, we introduce a dynamical Monte Carlo
algorithm for spin models \cite{NOTE}. 
In the proposed algorithm, the number of the spins (the system size)
fluctuates from zero to a given number 
by addition and deletion of spins with a probabilistic rule.
A key feature of the algorithm is the use of an artificial mixture 
of canonical ensembles each of which corresponds to a system with different number of the spins. 
The weight of each component of the mixture
is controlled by a penalty term for the system size
and  systematically tuned  
in a preliminary run in a way similar to that in multicanonical-type algorithms 
\cite{BN91,BC92,Lee}. 
In the measurement run, the system grows
and shrinks without violating the detailed balance condition
and the correct canonical averages are obtained
if we measure the physical quantities only when 
its size is equal to the prescribed maximum size.  
Our idea has many common features to 
other algorithms based on artificial extensions (or compositions) of 
the canonical ensembles, say, multicanonical algorithm \cite{BN91,BC92} or simulated tempering algorithm \cite{MP92}. The only difference is
that we use the mixture of the canonical ensembles with different system 
sizes instead of the one with different temperatures in
simulated tempering algorithm.

The proposed algorithm is viewed as a variant of the methods based on the step-by-step construction of an approximate ground state of a spin system. 
For example, consider an Ising model with random couplings.
If we add Ising spins sequentially to the system 
and choose the direction of a new spin to lower the 
energy in each step, we can construct a candidate of the ground state of system. At finite temperature, however,
the detailed balance condition is hardly satisfied, 
if the system is assumed to only grow and never diminish.
A method to preserve the detailed balance condition is to consider the addition and deletion of a spin as a Metropolis-type trial and  
allowing the system freely to grow and shrink in a probabilistic rule.
The bi-directional change of the system size is also useful
to avoid the trapping in metastable states.
The size of the system, however, can 
vary from zero to a desired size, only when 
proper values of the penalty for the size are given. 
Without a penalty for the system size, the system size 
will be stuck around an uncontrolled value.
At this point, we introduce the idea of multicanonical ensemble ---
learning proper values of the penalty in a preliminary run. 
This idea leads to the proposed algorithm.

Our main aim to develop the algorithm is 
to accelerate the slow relaxation
at low temperature, which is particularly serious
in Monte Carlo simulations of randomly frustrated system.
While fast relaxation is attained in the 
high temperature components of the extended ensemble 
in simulated tempering algorithm, we expect fast relaxation 
in the ``small system size'' components of the ensemble
in the present algorithm. In the latter part of the paper, 
we will show an example in which the proposed algorithm successfully 
facilitate the mixing of Markov chain in a spin glass problem.

Several works with related ideas are found in the literature.
Ensembles of polymers with variable length have been 
used in the study of lattice polymers \cite{Sokal95}. Specifically,
Grassberger \cite{G93} developed an algorithm in which 
the penalty to the length of the 
polymer is tuned in a preliminary run.
Wilding \cite{Wil95} explore the coexistence region of 
Lennard-Jones fluid with a modification of 
the multicanonical ensemble extended in a two-dimensional space spanned by
a pair of linear combinations of the number of the particles and the energy.
These studies are regarded as natural developments of the preceding works
on grand canonical ensembles of the corresponding systems. 
While the fluctuation of the system size is limited 
in grand canonical ensembles, 
it can be arbitrary large in the ensembles defined with 
tuned penalty terms.
The situation is somewhat different in our case, because 
grandcanonical methods are, in itself, not popular
for spin models.
On the other hand, Wong \cite{W95} gives an interesting suggestion on 
variable-system-size ensembles
in the context of statistical information processing. 
It is, however, a part of a discussion paper and no concrete 
implementation of his method, ``sequential buildup'', is reported
\cite{W95r}.

Here we implement the algorithm 
for the Sherrington Kirkpatrick model of spin glass and show that the idea is worked well in this case. Let us consider the energy function of the SK model \cite{SK,SG}:
\begin{equation}
\label{SK}
E(\{S_i\})= - \frac{1}{2} \sum_{i=1}^{N} \sum_{j=1}^{N}
J_{ij} S_i S_j, \quad S_i \in \{\pm 1\} {}_,
\end{equation}
\begin{equation}
\label{SKB}
P(\{S_i\})= \frac{\exp(-E(\{S_i\})/T)}{Z} {}_.
\end{equation}
The constant $N$ is the system size and $Z$ is the partition function.
The coupling constants $\{J_{ij}\}$ are mutually independent random samples from the Gaussian distribution with a variance $1/\sqrt{N}$ and a mean zero.

In this case, it is easy to define a variable-system-size ensemble.
Allowing the value $S_i=0$ to the spin variables $S_i$, 
the energy function eq.\ref{SK} is, in itself, viewed as the energy 
of the variable-system-size ensemble:  
\begin{equation}
E(\{S_i\})= - \frac{1}{2}
\sum_{i=1}^{N} \sum_{j=1}^{N} J_{ij} S_i S_j, \quad S_i \in \{+1,-1,0\} {}_.
\end{equation}
The number of spins with nonzero values
$n=\sum_i |S_i|$ is regarded as the size of a modified system.
When $n=N$, i.e., all of the variables $S_i$ take the value of $1$ or $-1$, 
the original system is recovered.

As is mentioned earlier, 
we need a proper penalty function $f$ of the size $n$,
if we want the system size $n$ to fluctuate between zero and the original size
$N$ of the system. We set
\begin{equation}
\label{modE}
E(\{S_i\})= - \frac{1}{2}
\sum_{i=1}^{N} \sum_{j=1}^{N} J_{ij} S_i S_j + f(n), \quad S_i
\in \{+1,-1,0\}
\end{equation}
and choose the adequate value of $f(n)$ at each $n$ through a preliminary simulation.
We use a method based on a histogram construction, which is similar to
that used in entropic sampling \cite{Lee} and estimate the
values of $f(n)$ that give nearly equal frequencies of $n$ 
in the interval $0 \leq n \leq N$. 
After the proper form of the penalty function $f$ is 
determined by the iterative procedure, we perform a run for the measurement of the statistics. If we measure the statistics only when $n=N$, we recover
the correct canonical averages.

Results from an experiment 
with a sample ($N=50$) is shown in Fig.1. 
We measure the statistics
 $\frac{1}{N^2}\sum_{(i,j)} \langle S_iS_j \rangle^2$, where the 
summation runs over all pairs of the spins (We measure them directly; the real replica technique \cite{Y} is not used here.). The results of eight runs 
with different initial conditions and different random number seeds 
are recorded at four temperatures for each algorithm described below.

In the first panel (a), the result with the proposed algorithm
is shown. A heat bath algorithm (Gibbs sampling algorithm) 
is used to simulate the modified system defined by eq.\ref{modE}.
At each step of the algorithm, 
a spin is randomly chosen and we substitute the state 
$S_i$ of the spin for one of the possible states $0$, $+1$ and $-1$
with a conditional probability defined by the energy eq.\ref{modE}.    
Here, we define $N$ iterations of the step as one Monte Carlo Step
(The maximum system size $N$ is used in the definition 
instead of $n$. Note that the value of $n$ varies in the simulation.).  
200000 MCS after the initial 1000 MCS 
is used for tuning the penalty $f(n)$.  
The values of $f(n)$ are updated in every
10000 MCS in this preliminary run. 
Then, samples from another 200000 MCS are used 
for the calculation of the statistics. A dotted line in the figure indicates
a corresponding result with exchange Monte Carlo (Metropolis-coupled chain)
algorithm \cite{ICOT,G91,G95,I93a,I93b,H96}. Good agreement of both results 
supports the validity of the proposed algorithm.

For the purpose of comparison, the results with conventional algorithms are
shown in the second and third panels (b,c). The result shown in the panel (b) 
correspond to a conventional heat bath algorithm.  The algorithm 
is similar to that in (a), but
the original energy eq.\ref{SK} is used instead of eq.\ref{modE}
and the value $0$ of $S_i$ is not allowed, i.e., the system size
$n$ is fixed to $N$.     
After the initial 200000 MCS is discarded,
samples from another 200000 MCS are used for the calculation of statistics.
Clearly, the variance of the results in (b) is much larger than that in (a).
 
In the experiment in the panel (c), 
we attached a simulated annealing procedure to the simulation in (b), i.e., 
we gradually lower the temperature in a part of the initial
200000 MCS. We start from $T=1.0$ and decrease 
the temperature with the rate $5 \times 10^{-6}$/MCS. After we reach the target
temperature, we keep it through the rest of the simulation.
The addition of the initial annealing phase 
will cause escape from the shallow metastable states around 
the irreverent local minima with higher energies.
Although the variances of the results are
considerably decreased in (c) compared with that in (b), they are still higher than that in (a). 
\begin{fullfigure}
\figureheight{10cm}
\caption{
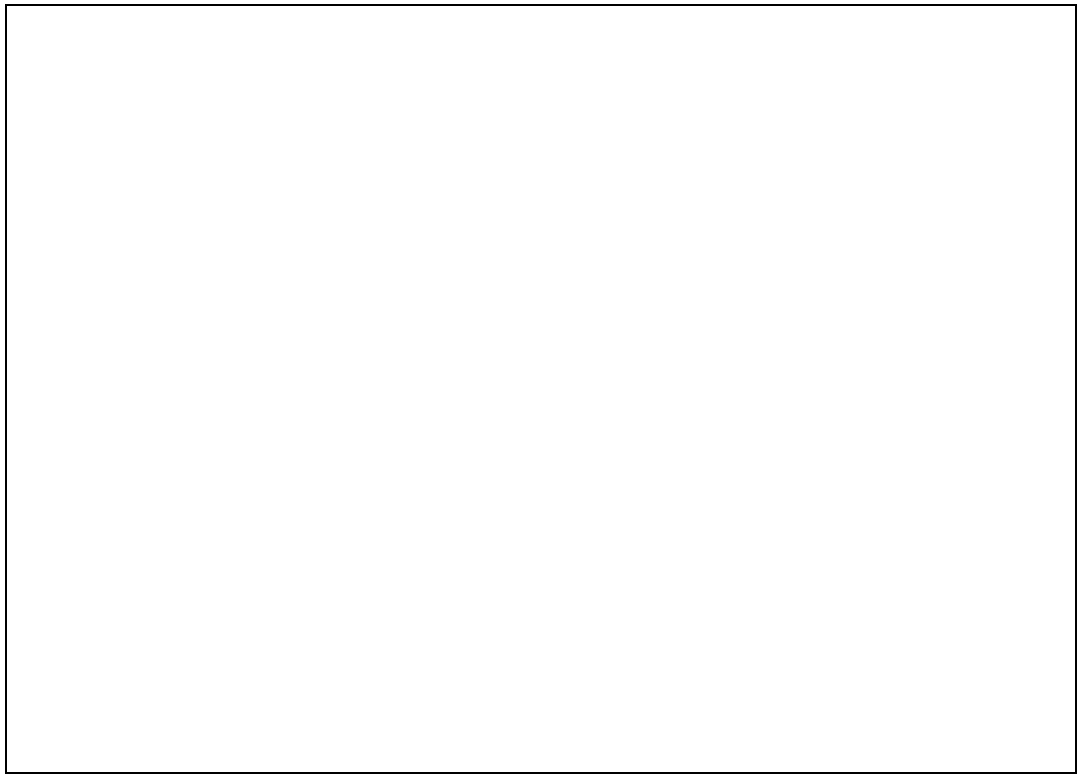
(a) The proposed algorithm. (b) A standard heat bath algorithm.
(c) A standard algorithm plus annealing (see text). 
The temperature $T$ and the statistics $\frac{1}{N^2}\sum_{(i,j)} \langle S_iS_j \rangle^2$
are shown in horizontal and vertical axes respectively.
In each panel, the results of eight independent runs
with different initial conditions and different random number seeds 
are shown (black dots). Dotted lines common 
in (a, b, c) show the values with a exchange Monte Carlo algorithm. 
}
\end{fullfigure}

In the proposed algorithm, we choose a spin randomly 
when we ``add'' or ``remove'' it. 
Then, there is a tendency that some spins in the system are
``likely to exist'' ($S_i \neq 0$) and others are not. As a result, the 
quenched disorder in the system
is no longer ``randomly generated'' one, when the system size $n$ is smaller
than the original size $N$ \cite{noteS}. This does not cause any bias of 
the calculated averages in our experiment because we discarded
the samples with $n \neq N$ from the measurement. Our experiment shows that 
we can get more independent samples with the proposed algorithm 
than with conventional algorithm, even with this loss of the samples.
On the other hand, we can design an algorithm where spins are added or removed 
with a prescribed order, e.g., $S_i$ can take the value $0$ only when
$S_ {i+1}=0$. With this algorithm, we can use the samples with $n < N$ as 
samples from a randomly generated system of the size $n$. Such a modification,
however, will lead to longer relaxation time and increases a risk of the bias of the averages.    

In this paper, we proposed a Monte Carlo algorithm for spin models with
a variable-system-size ensemble, a mixture of the canonical ensembles each of which corresponds to a modified system with 
different number of spins. The weight of each component of the mixture
is controlled by a penalty term for the system size and 
systematically tuned  in the preliminary run. The idea is successfully 
implemented for the SK model of spin glass, where the variable-system-size 
ensemble is easily realized with the introduction 
of the fictitious value $0$ of a spin that takes the values $\pm 1$
in the original model. We show that the proposed algorithm
shows a better performance than a conventional dynamical Monte 
Carlo algorithm in our example.

Although the success in this paper might partly rely on the property of the 
SK model, our method is fairly general and formally applicable to a number
of models, say, spin models on lattices or random networks. We can also
treat models with Potts or continuous spins. For the models with non-Ising 
spins, we should  introduce a set of labels, 
which shows that the corresponding spin
is assumed not to interact with any other spin and 
external field  when the label takes a prescribed ``kill'' value
\cite{noteA}.

At very low temperature, the proposed algorithm reduced to a greedy
algorithm to search ground states with step-by-step construction 
of the system and will be trapped
in a metastable state. In that case, a two-dimensional extension, e.g., 
a simultaneous extension in the energy and the system size
might be effective (See examples \cite{Wil95,SLV97,H97,ICK98} of
successful two-dimensional extensions.).


\begin{thebibliography}{9999}
\bibitem{noteS}
The study of the 
structure of the samples with $n<N$ might be interesting, 
if we can learn something about the original system from 
them.
\bibitem{noteA}
An additional rule is required 
to determine the state of a spin 
added to the system, when a heat bath type update
is computationally too expensive. It may 
cause some difficulty for the implementation
in continuous spin systems.
\end{thebibliography}
\end{document}